\documentclass[12pt]{article}
\usepackage{amssymb,amsmath,graphicx,latexsym}\pdfoutput=1
\begin{document}
\title{\textbf{Diquarks from a fourth family}} 
\author{B.~Holdom\thanks{bob.holdom@utoronto.ca}\\
\emph{\small Department of Physics, University of Toronto}\\[-1ex]
\emph{\small Toronto ON Canada M5S1A7}}
\date{}
\maketitle
\begin{abstract}
If fourth family condensates are responsible for electroweak symmetry breaking then they may also break approximate global symmetries. Among the resulting pseudo-Goldstone bosons are those that can have diquark quantum numbers. We describe the variety of diquarks and their decay modes, and we find aspects that are particular to the fourth family framework. Spectacular signatures at the LHC appear and are explored for color sextet diquarks with 600 GeV mass. We consider a simple search strategy which avoids diquark reconstruction. We also consider 350 GeV mass diquarks that are accessible at the Tevatron.
\end{abstract}

\section{Introduction}
Fourth family quarks with a mass in the range of the unitarity upper bound of 500 to 600 GeV will have strong couplings to the Goldstone bosons of electroweak symmetry breaking. It is then natural to expect that the dominant order parameters for electroweak symmetry breaking are in fact just the condensates of the fourth family fermions. If the underlying strong dynamics is a gauge theory then it must be a broken gauge symmetry, or else the fourth family picture becomes a technicolor picture. When the underlying gauge symmetry is broken then the heavy quarks can have standard CKM mixing with lighter quarks.

The heavy quark condensates generate the Goldstone bosons of electroweak symmetry breaking, but they can also generate pseudo-Goldstone bosons (PGBs) corresponding to approximate global symmetries of the underlying dynamics. This is true whether or not the strong gauge symmetry is itself breaking; all that matters are the approximate global symmetries that are being dynamically broken. For a discussion of these possible symmetries it is convenient to consider left-handed Weyl spinors, so that ($\psi_L$, $\psi^c_L$) expresses the content of a Dirac spinor where $\psi^c_L$ is the antiparticle of $\psi_R$. Now suppose that the strong interactions treat both $\psi_L$ and $\psi^c_L$ identically, that is they are in the same representation of the (broken) gauge symmetry. Then these interactions are invariant under global $SU(2)$ transformations acting on the ($\psi_L$, $\psi^c_L$) doublet. These transformations do not include the usual axial $U(1)$. If a normal Dirac mass condensate forms then this global $SU(2)$ is broken down to the vector $U(1)$. If the $SU(2)$ is only an approximate symmetry to begin with then the result is two PGBs carrying difermion quantum numbers, those of $\psi_L\psi_L$ and $\psi_R\psi_R$. When there is a whole fourth family of condensates then there will be variety of difermion PGBs with diquark, leptoquark and dilepton quantum numbers. Our focus here shall be on the diquarks and to a lesser extent the leptoquarks.

Although there may be other possibilities, the emergence of diquarks happens most simply if the strong gauge interaction is abelian. We label the (broken) gauge symmetry as $U(1)_X$. We take it to be a remnant of a larger flavor gauge symmetry that is mostly broken at a higher scale, perhaps around 1000 TeV. The $X$ boson remnant has a mass closer to 1 TeV. Its couplings to fermions should respect a custodial symmetry and it should be anomaly free. It is simple and natural to cancel anomalies by taking $X$ charges to be equal and opposite for the third and fourth families. The third family couplings of such a gauge boson has implications at the LHC which have been considered elsewhere \cite{a1}.

Of interest to us here is that there are naturally fields of opposite quark number that carry the same $X$ charge. These approximate symmetries are inevitably broken by the condensates and diquark PGBs are the result. We shall explore the interplay between the properties of the diquarks and the choice of $X$ charges of the third and fourth family quarks. Arranging the fields as $(q'_L,q'_R,q_L,q_R)$ with $q'=(t',b')$ and $q=(t,b)$, we shall consider two possible $X$ charge assignments: ${\cal Q}_A: (+,-,-,+)$ and ${\cal Q}_V: (+,+,-,-)$. We describe the resulting diquarks and also briefly the leptoquarks. We then turn to the pair production of diquarks at hadron colliders where we shall narrow our focus further to the color sextet diquarks.

\section{The $X$ charges of $(q'_L,q'_R,q_L,q_R)$}
\subsection{${\cal Q}_A: (+,-,-,+)$ }
Here the fourth family quarks $q'_L$ and ${q'}^c_L$ have the same $X$ charge and so this is the case mentioned above. $q'_L$ and ${q'}^c_L$ each represent six fields due to color and isospin. The previous $SU(2)$ symmetry now becomes a $SU(12)$ symmetry of the ($q'_L$, ${q'}^c_L$) fields. The fourth family quark condensates $\langle \overline{q'}q'\rangle$ break this down to $SO(12)$.

In the basis ($q'_L$, ${q'}^c_L$) these condensates are proportional to the $12\times12$ symmetric matrix
\begin{equation}
M=\left(\begin{array}{cc}0 & I \\I & 0\end{array}\right)
\end{equation}
where $I$ is the $6\times6$ identity matrix. An infinitesimal $SU(12)$ transformation produces a change proportional to $Y^T M+MY$ where $Y$ is a $SU(12)$ generator. In this basis we can take the 66 unbroken generators to be
\begin{equation}
\left(\begin{array}{cc}I & 0 \\0 & -I\end{array}\right)\quad
\left(\begin{array}{cc}T & 0 \\0 & -T\end{array}\right)\quad
\left(\begin{array}{cc}0 & A_1 \\0 & 0\end{array}\right)\quad
\left(\begin{array}{cc}0 & 0 \\A_2 & 0\end{array}\right)
\end{equation}
where $T$ and $A_{1,2}$ are hermitian and antisymmetric respectively. The broken generators are
\begin{equation}
\left(\begin{array}{cc}T & 0 \\0 & T\end{array}\right)\quad
\left(\begin{array}{cc}0 & S_1 \\0 & 0\end{array}\right)\quad
\left(\begin{array}{cc}0 & 0 \\S_2 & 0\end{array}\right)
\end{equation}
where $S_{1,2}$ is symmetric. The $S_{1,2}$ generators produce a change in $M$ that can be seen to correspond to the diquark PGBs. There are 42 of these made up of 6 color sextets and 2 color triplets. The $SU(3)_C\times SU(2)_L\times U(1)_Y$ quantum numbers of the sextets are (6,3,1/3) for $t'_Lt'_L$, $[t'_Lb'_L]$, $b'_Lb'_L$ and (6,1,4/3), (6,1,1/3), (6,1,-2/3) for $t'_Rt'_R$, $[t'_Rb'_R]$, $b'_Rb'_R$, while for the triplets they are ($\overline{3}$,1,1/3) for $\{t'_Lb'_L\}$ and $\{t'_Rb'_R\}$. [] and \{\} denote symmetric and anti-symmetric combinations. If we label the diquarks in terms of electric charge and color, there are two each of $\Phi_6^{4/3}$, $\Phi_6^{1/3}$, $\Phi_6^{-2/3}$ and $\Phi_{\overline{3}}^{1/3}$.

Of the remaining 35 PGBs there are four color octets (an electroweak triplet and a singlet) and a color singlet electroweak triplet. The latter in combination with the leptonic analogs form the three Goldstone bosons of EWSB along with three PGBs. Most PGBs receive mass from standard model gauge interactions and all can receive mass from other flavor physics effects arising at higher scales. 

\subsection{${\cal Q}_V: (+,+,-,-)$}
Now the Weyl spinors that have identical $X$ charge belong to different families. We arrange the fields as ($q'_L$, ${q}^c_L$, $q_L$, ${q'}^c_L$) so that the $X$ charges are $(+,+,-,-)$. Then a $SU(12)_1\times SU(12)_2$ symmetry is represented by block diagonal matrices
\begin{equation}
\left(\begin{array}{cc}Y_1 & 0 \\0 & Y_2\end{array}\right)
\end{equation}
in a 24-dimensional space. In this space the fourth family condensates are proportional to
\begin{equation}
M=\left(\begin{array}{cccc}0 & 0 & 0 & I \\0 & 0 & 0 & 0 \\0 & 0 & 0 & 0 \\I & 0 & 0 & 0\end{array}\right)
\end{equation}
where $I$ is again $6\times6$. Now the broken generators are
\begin{equation}
\left(\begin{array}{cccc}T & 0 & 0 & 0 \\0 & 0 & 0 & 0 \\0 & 0 & 0 & 0 \\0 & 0 & 0 & T\end{array}\right)\quad
\left(\begin{array}{cccc}0 & V_1 & 0 & 0 \\0 & 0 & 0 & 0 \\0 & 0 & 0 & 0 \\0 & 0 & 0 & 0\end{array}\right)\quad
\left(\begin{array}{cccc}0 & 0 & 0 & 0 \\0 & 0 & 0 & 0 \\0 & 0 & 0 & 0 \\0 & 0 & V_2 & 0\end{array}\right)
\end{equation}
The $V_{1,2}$ have both symmetric and antisymmetric parts where the symmetric parts give the same set of 42 diquarks as before. The antisymmetric parts give rise to an additional 30 diquarks, 2 color sextets and 6 color triplets. Their $SU(3)_C\times SU(2)_L\times U(1)_Y$ quantum numbers are (6,1,1/3) for the two sextets and ($\overline{3}$,1,4/3), ($\overline{3}$,1,1/3), ($\overline{3}$,1,-2/3) and ($\overline{3}$,3,1/3) for the triplets. For this $X$ charge assignment all 72 diquarks have a flavor content consisting of one fourth family quark and one third family quark.

${\cal Q}_A$ is not gauged and instead specifies an approximate global symmetry that is broken by the condensates.  Thus there is one additional color singlet, neutral PGB. It is of interest that the ${\cal Q}_A$ approximate symmetry is consistent with a large top mass; there is an operator that can generate a top mass from a fourth family condensate, $\overline{b}'_Lb'_R\overline{t}_L t_R$, which preserves both ${\cal Q}_A$ and ${\cal Q}_V$ charge \cite{a7}. (The operator $\overline{t}'_Lt'_R\overline{t}_R t_L$ preserves ${\cal Q}_V$ but not ${\cal Q}_A$.) For this reason this PGB could be the lightest of the possible color and isospin singlet PGBs. The couplings of this PGB to the fourth family implies that it has loop-induced couplings to $gg$, $\gamma\gamma$, $ZZ$ and $WW$.

For this ${\cal Q}_V$ charge assignment the $U(1)_X$ is vectorial with respect to the fourth family quark condensates, and so the latter do not break the $U(1)_X$ gauge symmetry. Something else must produce the $X$ mass and this could include leptonic condensates and/or four-fermion condensates. Finally we see that the quark condensates are occurring in the naively attractive channel (from one $X$ boson exchange), in contrast to the ${\cal Q}_A$ charge assignment. But we don't view the one gauge boson exchange prediction as being very trustworthy when $U(1)_X$ is strongly interacting and/or when there are other possible effects that can influence symmetry breaking.

\section{Discussion}
We can compare the diquark content in the previous two cases to the ditechniquark technipions that arise in a technicolor theory when the technifermions are in real technicolor representations (e.g. see \cite{a3}). When the invariant two index symbol of the technicolor group is symmetric the set of ditechniquark technipions matches the set of diquarks in the case of the ${\cal Q}_A$ charge assignment. The additional set of diquarks for the ${\cal Q}_V$ charge assignment corresponds to the set of ditechniquark technipions occurring when the invariant symbol is antisymmetric. The simultaneous existence of both sets of diquarks thus differs from the ditechniquark content of technicolor theories.

The diquark masses are model dependent but to have any meaning as PGBs the masses should be less than the sum of the masses of the quark constituents. The QCD contribution puts a lower bound on the diquark masses and this can be estimated as is done for colored technipions, by scaling up the electromagnetic mass difference of pions and accounting for color factors \cite{a3}. This puts the QCD contribution at 
\begin{equation}
M_{\Phi}^{\rm QCD}\approx\left(\frac{C_2(R)\alpha_c}{\alpha}\right)^{1/2}\frac{F_4}{f_\pi}\;35.5\mbox{ MeV}
\end{equation}
where $F_4\approx250/\sqrt{N_d}$ GeV. The number of doublets $N_d$ is effectively between 3 and 4 since the fourth family leptons are expected to contribute less than one third of a quark doublet. This puts the QCD contribution to sextet and triplet masses at very roughly 350 and 220 GeV respectively. Flavor physics, in particular associated with the top mass, will also break diquark symmetries and this will yield additional contributions to the diquark masses. Given that $2m_{q'}$ could be around a TeV we are left with a wide range of possible diquark masses.

Diquarks have recently attracted attention in the literature since some of them can be fairly light and have substantial couplings to light quarks and yet still evade limits on flavor changing processes \cite{a6}. But it is extremely model dependent as to whether diquarks could actually emerge with light quark couplings large enough to produce interesting effects, such as being produced singly \cite{a9} at hadron colliders with observable cross sections. Instead we shall focus on the pair production of diquarks through their gluonic coupling, which is completely determined by the color representation of the diquark.

The decay products of a diquark must include two fermions, and as we shall see, $W$'s may also be produced. These decays can produce quite spectacular events when these fermions are $t$'s and/or $b$'s. (In particular the color octet PGBs are less interesting since they decay to two gluons or to a gluon and a weak gauge boson or to a gluon and a color singlet PGB.) We first consider the direct decay of a diquark to two fermions. The required couplings between the diquarks and the lighter fermions can be induced by effective four fermion operators. These couplings are thus very model dependent and there is no reason to expect a proportionality to the mass of the lighter fermion. In fact the relevant four fermion operators are quite unrelated to those that generate mass.

For the case of the ${\cal Q}_A$ charge assignment, a coupling of a diquark to lighter quarks would have to arise from operators like $\overline{q}_{iL}\gamma_\mu q'_L\overline{q}_{jL}\gamma^\mu q'_L$ or $\overline{q}_{iR} q'_L\overline{q}_{jR} q'_L$ and the same with $(L\leftrightarrow R)$ where $i,j=1..3$ is a family index. For the ${\cal Q}_V$ charge assignment one of the $q'$'s would need to be replaced by a $q_3$. These operators must be $SU(3)_C\times SU(2)_L\times U(1)_Y$ symmetric but other than that the light  flavor structure of these operators is model dependent. These operators are the analogs of flavor changing $\Delta F=2$ operators among lighter quarks. If they are similarly suppressed then the diquark couplings to lighter fermions could be much smaller than the size of standard Yukawa couplings. We also note that the chirality changing operators such as $\overline{q}_{iR} q'_L\overline{q}_{jR} q'_L$ can only couple $SU(2)_L$ singlet, charge $1/3$ diquarks to lighter quarks.  

Besides the direct decay to two fermions, the diquarks of a fourth family can also decay weakly. This occurs as long as the heavy quarks themselves decay weakly, that is as long as there is some CKM mixing with the lighter families. These are decay modes not shared by the ditechniquarks of a technicolor theory. For the ${\cal Q}_A$ charge assignment a diquark can decay through a virtual pair of the heavy quarks.  For now we assume that the heavy quarks decay to third family quarks. Then for the sextet diquarks we have
$$\Phi_6^{-2/3}\rightarrow b'^*b'^*\rightarrow ttW^-W^-$$
$$\Phi_6^{1/3}\rightarrow t'^*b'^*\rightarrow btW^+W^-$$
\begin{equation}
\Phi_6^{4/3}\rightarrow t'^*t'^*\rightarrow bbW^+W^+
.\label{e1}\end{equation}
All these decays can be treated as occurring through effective operators obtained by integrating out the heavy quarks. In the case of $\Phi_6^{1/3}$ decay the external $W^+W^-$ can be replaced by an internal $W$ exchange, and so this produces an additional contribution to the direct decays of charge $1/3$ diquarks.

For the ${\cal Q}_V$ charge assignment there is only one heavy quark in the decays, and this is a virtual quark if the diquark mass is below the 2-body decay threshold.
$$\Phi_6^{1/3}\rightarrow b'^*t\rightarrow ttW^-$$
$$\Phi_6^{-2/3}\rightarrow b'^*b\rightarrow tbW^-$$
$$\Phi_6^{4/3}\rightarrow t'^*t\rightarrow btW^+$$
\begin{equation}
\Phi_6^{1/3}\rightarrow t'^*b\rightarrow bbW^+
\label{e2}\end{equation}

We note that the various types of decays can violate the ${\cal Q}_A$ and ${\cal Q}_V$ charges. With the ${\cal Q}_A$ charge assignment the diquarks carry two units (in absolute value) of both charges. Decays to two third family quarks of the chirality preserving (or changing) type violates both charges (or only the ${\cal Q}_V$ charge). With the ${\cal Q}_V$ charge assignment the diquarks have zero ${\cal Q}_A$ and ${\cal Q}_V$ charges, and so decays to two third family quarks violates both charges. This discussion also applies to the weak decays since CKM mixing between the third and fourth families also violates both charges. These approximate symmetries thus suggest that all these decay modes are suppressed while leaving open the question of whether it is the direct or weak decays that dominate. They also make it possible that decays involving first or second family quarks could be important, since perhaps some extended approximate symmetry is preserved in that case.

\subsection{Leptoquarks}
Since leptoquarks are also of interest at hadron colliders we briefly describe them. They are all color triplets and so we need only give their flavor content, as obtained by extending the treatment above. We only consider the leptoquarks with fermion number two. For the ${\cal Q}_A$ charge assignment carried over to leptons this content is $t'\nu'_\tau$, $t'\tau'$, $b'\nu'_\tau$, $b'\tau'$ while for the ${\cal Q}_V$ charge assignment it is $t'\nu_\tau$, $t'\tau$, $b'\nu_\tau$, $b'\tau$, $t\nu'_\tau$, $t\tau'$, $b\nu'_\tau$, $b\tau'$. For each flavor content there is a pair of leptoquarks since the fermions can be left- or right-handed. If the right-handed neutrinos don't exist then the leptoquarks containing them don't exist. The non-existence of the right-handed neutrinos also implies that the left-handed $\nu'_\tau$ has a Majorana condensate.\footnote{The beneficial impact of such a condensate on electroweak corrections has been described in \cite{a8}. This condensate would also contribute to the $X$ mass for either charge assignment.} In this case the leptoquarks containing left-handed neutrinos remain the same as before except for the ${\cal Q}_V$ charge assignment where they become $t'\nu_\tau$, $b'\nu_\tau$, $t'\nu'_\tau$, $b'\nu'_\tau$. Since the $\tau'$ could be significantly heavier than the $\nu'_\tau$, the plausible decays of the fourth family leptons are $\nu'_\tau\rightarrow\ell W^+$ ($\ell=e$, $\mu$, or $\tau$) and $\tau'\rightarrow\nu'_\tau W^-\rightarrow\ell W^+W^-$. Then the weak decay of a leptoquark can produce a $t$ or $b$ plus a charged lepton plus up to three $W$'s depending on the leptoquark.

\section{Pair production at hadron colliders}
\subsection{LHC}
When produced from a $q\overline{q}$ initial state the pair production of diquarks suffers from the usual $p$-wave suppression. But this does not hinder their production at the LHC where their production from the $gg$ initial state escapes this suppression. For this initial state we also observe that the color sextet cross section is about 20 times as large as for the color triplet (from Madgraph \cite{C}). (For a $q\overline{q}$ initial state it is 5 times as large.) This is the main reason we choose to focus on the sextets rather than the triplet diquarks or the leptoquarks.  We first consider sextet diquark masses of 600 GeV. Their cross section for $\sqrt{s}=7$ TeV at the LHC is about 40\% larger than for 600 GeV quarks. The number of different sextets (six or eight depending on the $X$ charge assignment)  boosts the total cross section into diquarks even further.

One obvious signature is same sign leptons and this was investigated for the $tt\overline{t}\overline{t}$ signal from $\Phi_6^{4/3}\rightarrow tt$ in \cite{a2}. If both of these $t$'s decay semileptonically to produce the same sign signal then reconstruction of both hadronically decaying $\overline{t}$'s from the $\overline{\Phi}_6^{4/3}$ can be attempted as well \cite{a2}. We do not pursue the same sign lepton signal or diquark reconstruction further here.

Many jets, $b$-jets, isolated leptons and missing energy will characterize diquark pair production, especially when $W$'s are also produced. A particular feature of these events could be useful for a first detection of a signal above background; this is the large number of energetic objects and the relatively uniform partition of energy among these objects. A simple diagnostic for this is obtained from the two objects (jets or otherwise) in each event having the highest transverse energies. The point is to compare the sum of these two energies with the total transverse energy $H_T$ and so we define
\begin{equation}
H_2\equiv\frac{E_{1T}+E_{2T}}{H_T}
.\label{e3}\end{equation}
We shall see that $H_2$ is typically less than 0.4 for the signal events while it is typically more than 0.4 for the background events.

The background can be reduced significantly relative to the signal with the following cuts: 1) an $H_T$ cut close to the peak of the $H_T$ distribution of the signal, which is about 1 TeV for a 600 GeV diquark; 2) at least six jets with $p_T>30$ GeV; 3) at least one isolated lepton with $p_T>15$ GeV; 4) at least one $b$-tagged jet; 5) missing energy greater than 20 GeV. The main remaining background with these cuts is due to $t\overline{t}+\mbox{jets}$ production.
\begin{center}\includegraphics[scale=0.5]{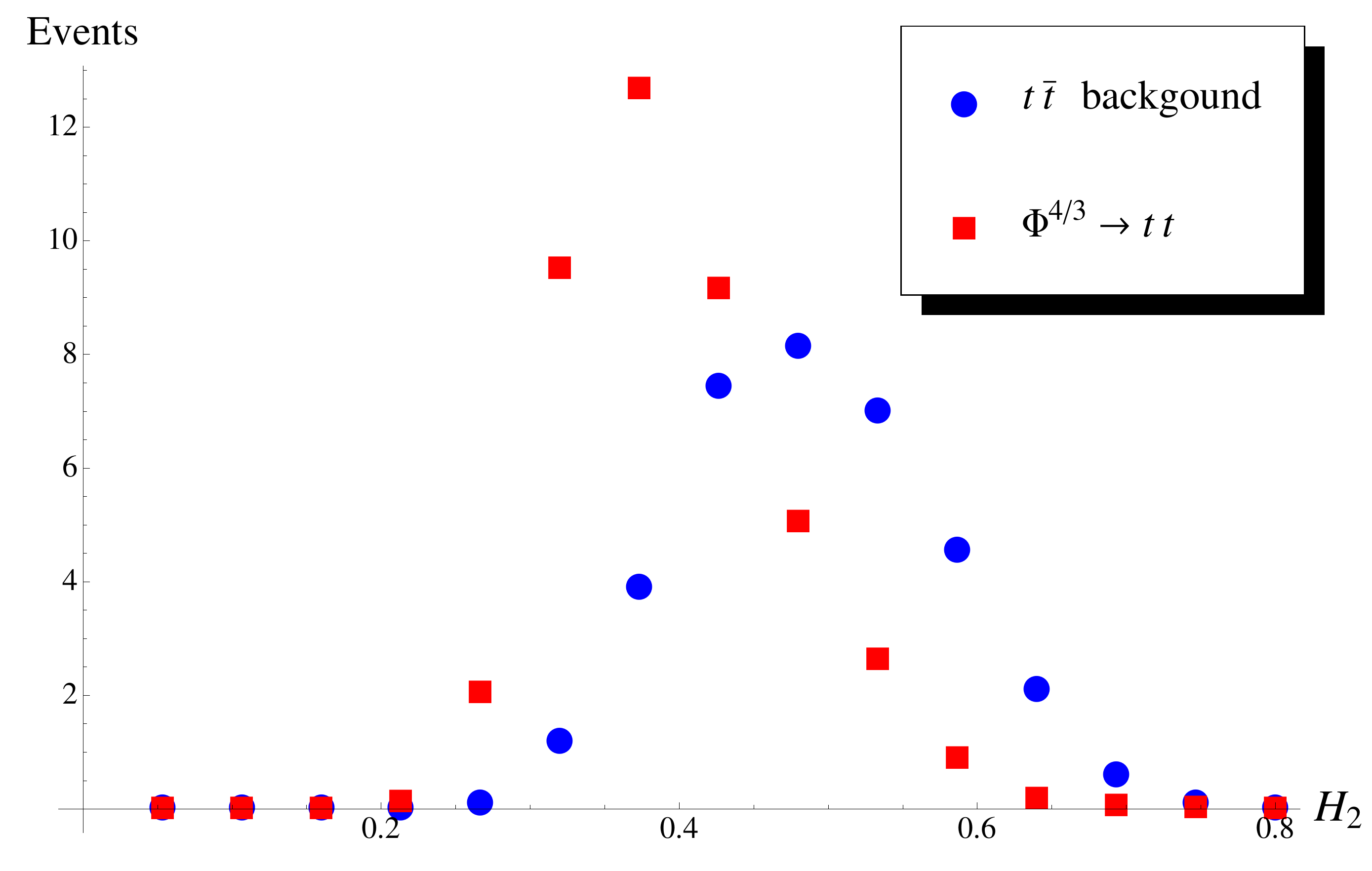}\end{center}
\vspace{-3ex}\noindent Figure 1: Number of events for 1 fm$^{-1}$, $\sqrt{s}=7$ TeV, and $H_2$ defined in (\ref{e3}).
\begin{center}\includegraphics[scale=0.5]{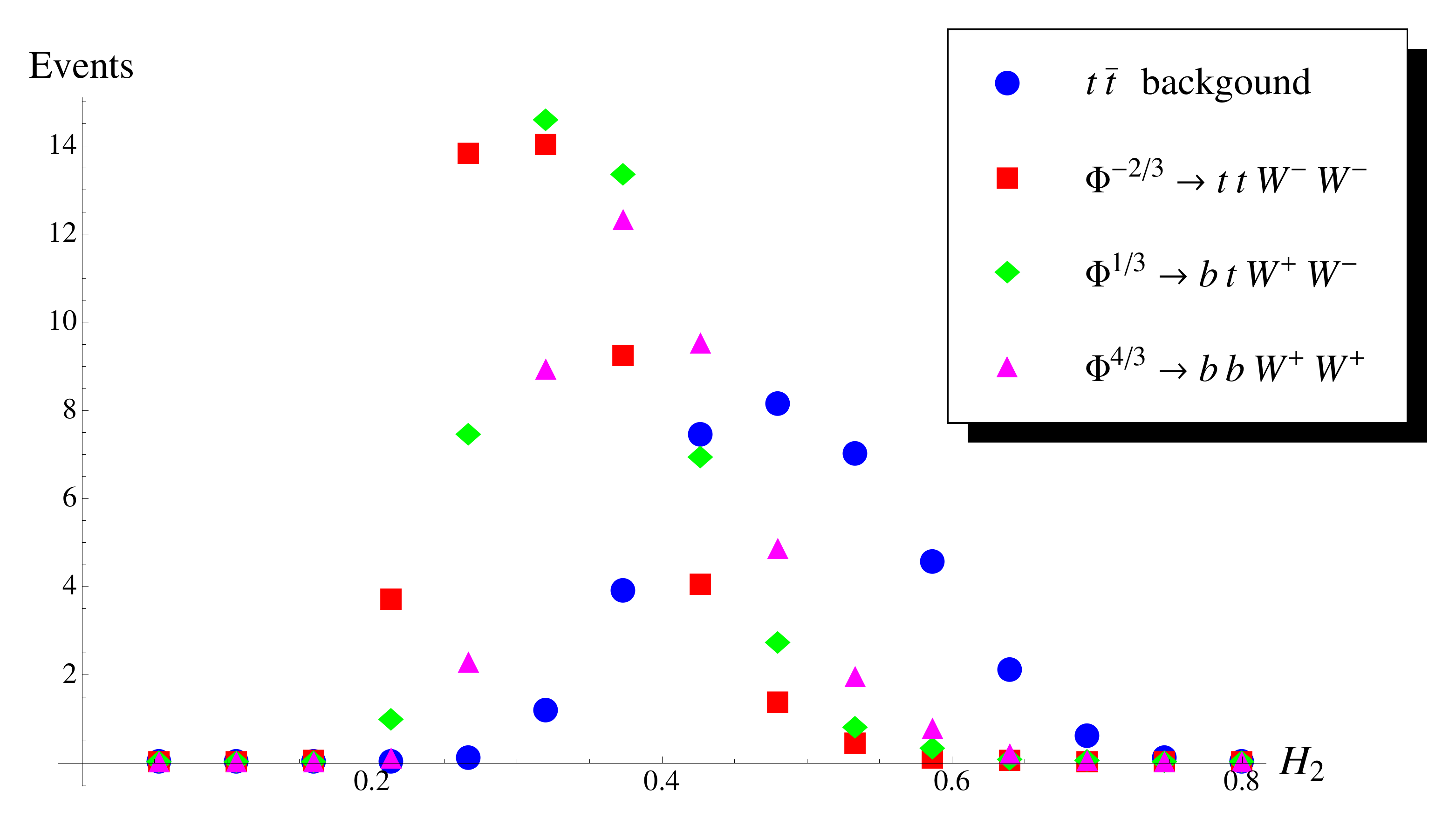}\end{center}
\vspace{-3ex}\noindent Figure 2: Number of events for 1 fm$^{-1}$, $\sqrt{s}=7$ TeV, and $H_2$ defined in (\ref{e3}).
\begin{center}\includegraphics[scale=0.5]{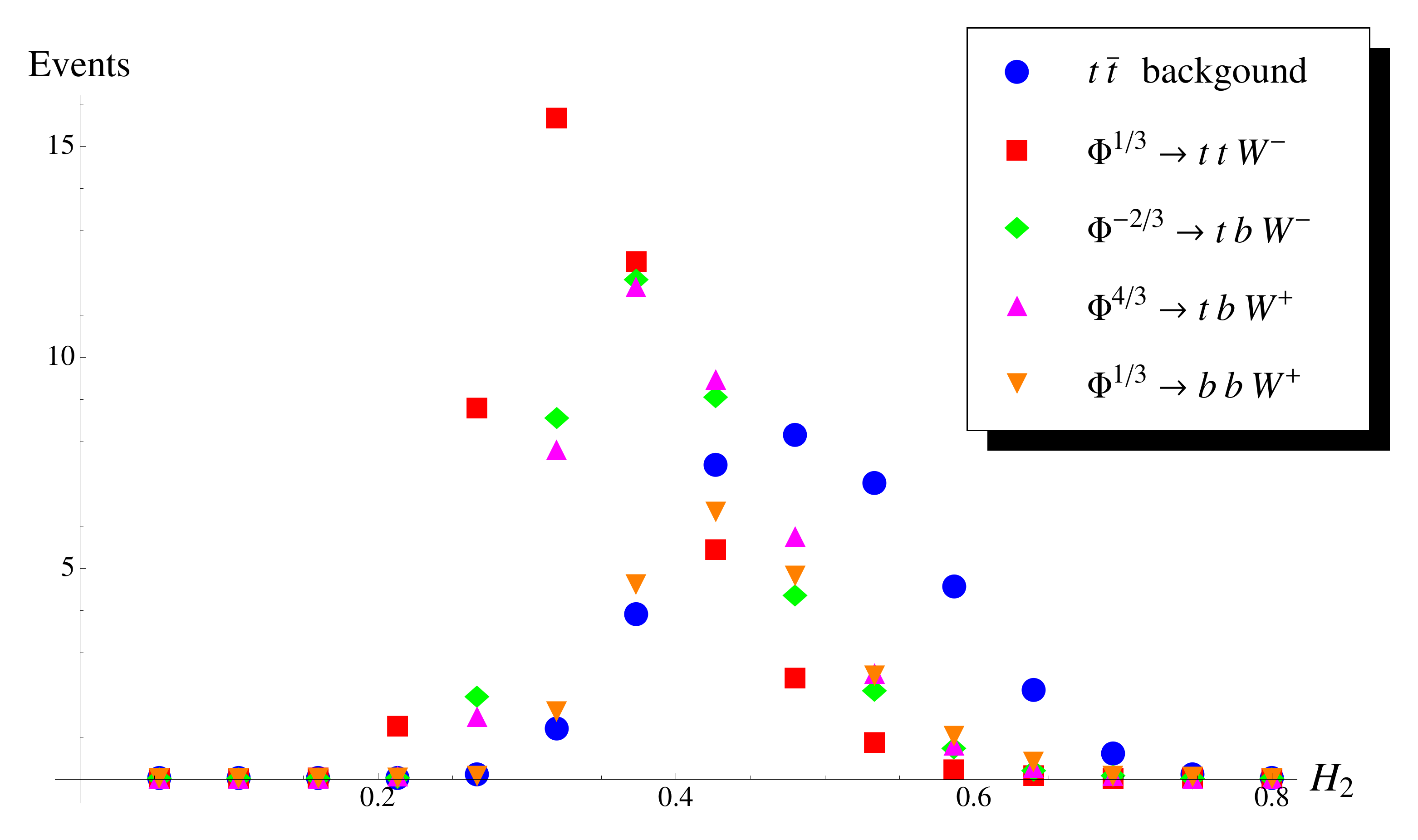}\end{center}
\vspace{-3ex}\noindent Figure 3: Number of events for 1 fm$^{-1}$, $\sqrt{s}=7$ TeV, and $H_2$ defined in (\ref{e3}).
\vspace{2ex}

In the figures we display the distributions in $H_2$ for events passing these cuts, with the event numbers in each case corresponding to 1 fm$^{-1}$. Fig.~(1) shows the case $\Phi_6^{4/3}\rightarrow tt$, the most interesting of the direct decays, along with the $t\overline{t}+\mbox{jets}$ background. The decays in (\ref{e1}) and (\ref{e2}) are shown in Figs.~(2) and (3) respectively. The cross sections have been doubled since there are two states (with left- or right-handed quark content) for each of the diquarks listed. The diquark decay widths could be quite narrow and we need only assume that they decay within the detector. We are assuming that they have a 100\% branching fraction into the given decay modes.

We see from these figures that a cut at $H_2<0.4$ will substantially increase signal to background.  The fraction of background events with $H_2<0.4$ is 21\% while this fraction is 87\%, 77\% and 58\% for the three processes in (\ref{e1}) and 81\%, 58\%, 53\% and 29\% for the four processes in (\ref{e2}). For $\Phi_6^{4/3}\rightarrow tt$ the ratio is 58\%.

We have mentioned that it is conceivable that the heavy quarks prefer to decay to quarks of the first two families. Then the events are somewhat less spectacular. For the decays in (\ref{e1}) the $H_2$ distributions for all four decays would be the same as what we have shown for $\Phi_6^{4/3}\rightarrow bbW^+W^+$. The $b$'s would be replaced by light quarks and so $b$-tagging would no longer be effective. For the decays in (\ref{e2}), two of the $H_2$ distributions would look like the $\Phi_6^{4/3}\rightarrow btW^+$ case and the other two would look like the $\Phi_6^{1/3}\rightarrow bbW^+$ case. Just one $b$ in each of these cases would be replaced by a light quark.

We note that the situation is opposite for the leptoquarks. Should their decays involve $e$'s or $\mu$'s rather than $\tau$'s then this will produce more striking signals.

\subsection{Tevatron}
We now consider lighter sextet diquarks with a mass of 350 GeV, which is close to the QCD contribution to their mass. This makes them accessible to the Tevatron and so we perform the analysis in that context. Now the $q\overline{q}$ initial state dominates and this brings in a $p$-wave suppression.

The same cuts are used as before except that the $H_T$ cut is reduced to 550 GeV and the six jets are now required to have $p_T>20$ GeV. The distributions in $H_2$ are shown in Figs.~(4) and (5) with the event numbers now corresponding to 10 fm$^{-1}$. We only consider those processes in (\ref{e1}) and (\ref{e2}) that have non-negligible phase space for decay and we have added $\Phi_6^{1/3}\rightarrow tb$ in case the direct decay dominates. Again we see that the decays producing more final state particles are differentiated from background with the simple $H_2$ diagnostic.

The window of opportunity for the Tevatron to find light diquarks before the LHC may be small or nonexistent. For a 350 GeV diquark, 1 fb$^{-1}$ of integrated luminosity at the LHC can produce about 40 times more signal events than produced with 10 fb$^{-1}$ at the Tevatron. In this comparison the previous LHC cuts are used but with $H_T>550$ GeV. We find that the $H_2$ diagnostic is less useful for 350 GeV diquarks at the LHC than it was for 600 GeV diquarks at the LHC or 350 GeV diquarks at the Tevatron. But 350 GeV diquarks at the LHC should show up by simply counting an excess of high activity events passing the cuts we have described and by employing a scan over the choice of $H_T$ cut.

\begin{center}\includegraphics[scale=0.5]{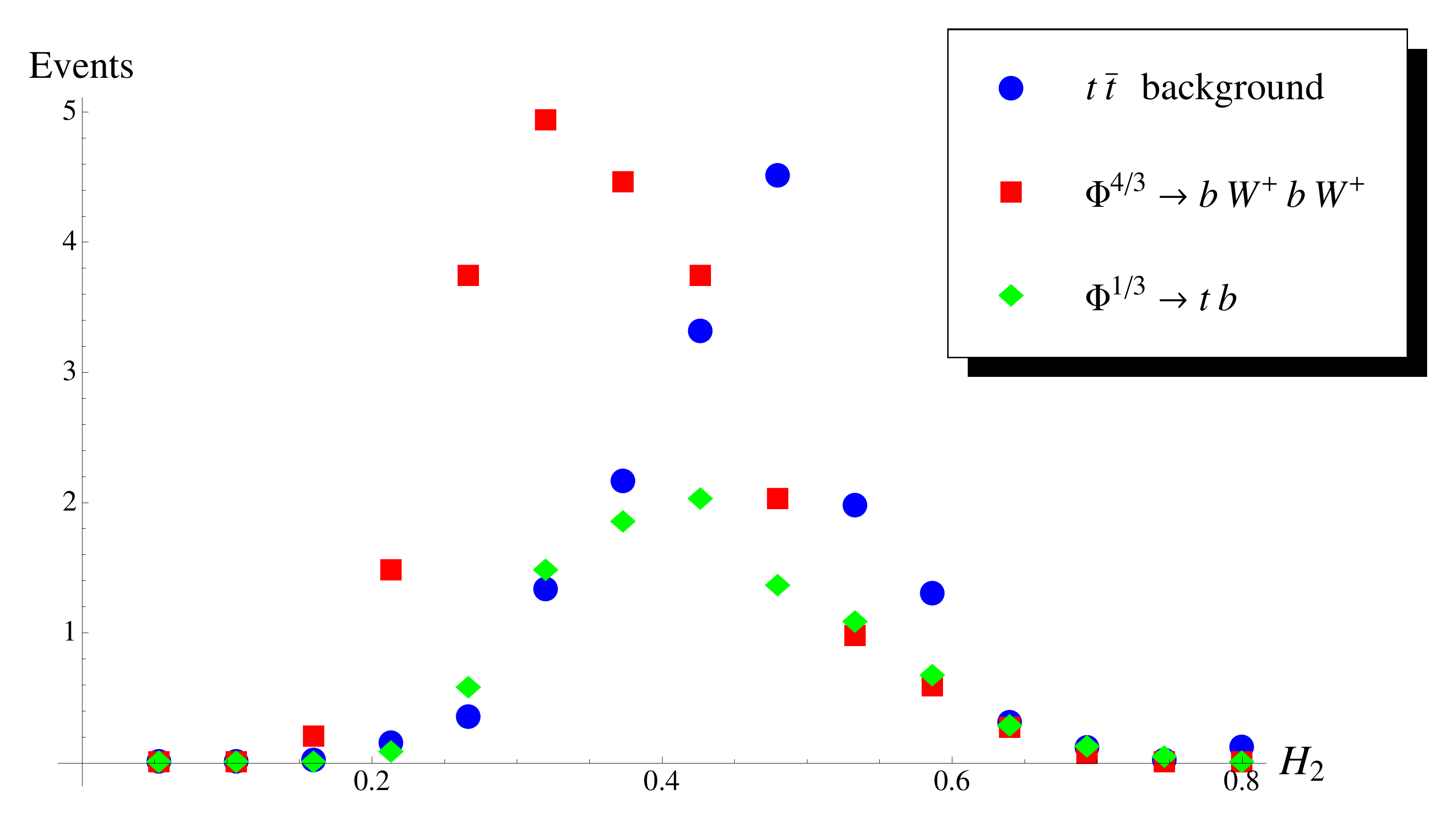}\end{center}
\vspace{-3ex}\noindent Figure 4: Number of events for 10 fm$^{-1}$ at the Tevatron.
\begin{center}\includegraphics[scale=0.5]{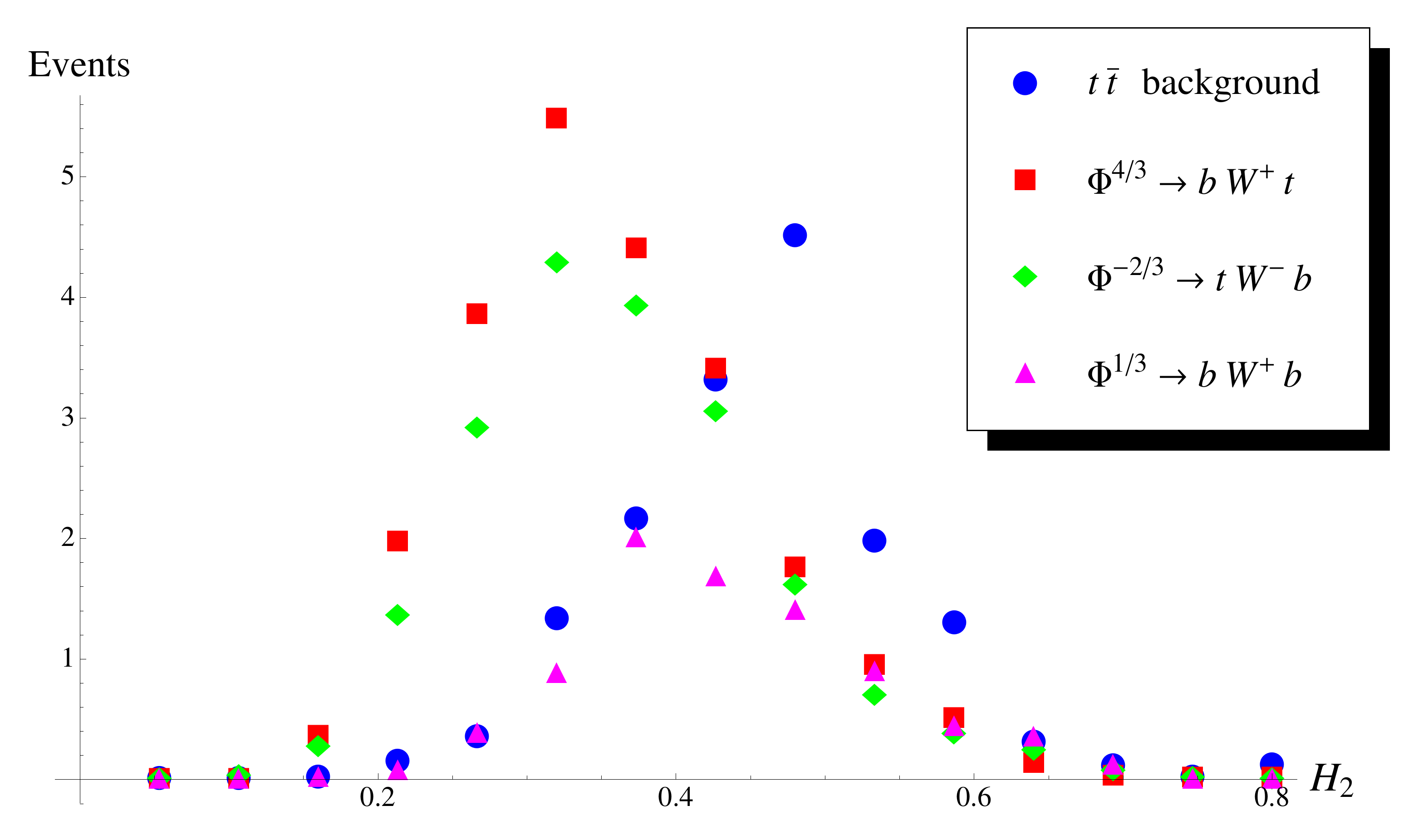}\end{center}
\vspace{-3ex}\noindent Figure 5: Number of events for 10 fm$^{-1}$ at the Tevatron.
\vspace{2ex}

\subsection{Event generation details}
We have used Madgraph 5 \cite{C} along with FeynRules \cite{E} to generate the required vertices for Madgraph. We combined the available FeynRules models \cite{E} for a fourth family and for a sextet diquark. The diquark decays in (\ref{e1}) are effectively four body decays and we can fake such decays via two body decays if we assign a fake mass and width to what corresponds to the virtual heavy quark. The point is to bring the mass down to allow a two body decay but at the same time make this intermediate quark state very broad. In particular we use a mass $1/2$ the diquark mass and a width equal to the diquark mass. Then the distribution of invariant masses of the $Wq$ emerging from the ``virtual quark'' will be fairly flat over the kinematically allowed range, as would be expected for the true invariant mass distributions of the 4-body phase space.  The decays in  (\ref{e2}) are handled similarly.

The $t\overline{t}+\mbox{jets}$ background was generated with Alpgen \cite{B} and MLM parton-jet matching. We used $b$ tagging efficiencies of 0.6, 0.1 and 0.01 for $b$, $c$ and light quarks respectively. Pythia \cite{A} and PGS \cite{D} were used for parton showering and detector simulation.

\section{Summary}
In a picture where fourth family condensates are responsible for electroweak symmetry breaking we have discussed the appearance of diquark pseudo-Goldstone bosons. The approximate symmetries required for such PGBs can appear as follows. Some remnant of a broken flavor gauge symmetry survives down to TeV scales and since it is strong it plays at least a partial role in inducing the condensates. There are two reasons why this $X$ boson couples to the third family as well as the fourth. One is that this allows for a simple cancellation of gauge anomalies. The other is that such an interaction can enhance the operator responsible for the top mass via an anomalous scaling \cite{a7}. If the $X$ charges of the various quark fields have the same absolute value, then because of anomaly cancellation there must be Weyl spinors that have opposite quark number with the same $X$ charge. Then the strong interaction has the global symmetries required for diquarks, and such symmetries will be broken no matter how the fields decide to pair up to form the fourth family condensates.

We have considered two possibilities for this pairing such that the resulting diquarks are either composed entirely of the fourth family quarks or they are composed of one fourth and one third family quark. The latter case yields diquarks that are both symmetric and antisymmetric in the combined color-flavor space of two quarks. The decays of the diquarks also reflect their fourth family origin, since the fourth family quarks can themselves decay weakly. This leads to diquark decays into two quarks along with one or two $W$'s. When these quarks are $t$'s and $b$'s then there are spectacular events with many distinguishing features (many jets, $b$-jets, leptons, missing energy). We have introduced a simple diagnostic to further distinguish these events from background.

The color sextet diquark signals we have presented should make them at least as easy to find as heavy quarks of similar mass. If heavy quarks are found but not the diquarks then either the diquark masses are larger and/or they have decays that are less spectacular, such as when direct decays to light quarks dominate. Alternatively their non-observation could mean that a gauge interaction as we have described it does not play a major role in the generation of the heavy quark condensates.

\section*{Acknowledgments}
This work was supported in part by the Natural Science and Engineering Research Council of Canada.

\end{document}